\documentclass[aip,reprint,floatfix,onecolumn]{revtex4-1}
\usepackage{amsfonts,amssymb,amsmath,array}

\usepackage{bm}

\usepackage[final]{graphicx}
\usepackage{graphicx}
\usepackage{epstopdf}
\graphicspath{{./Figures/}} 

\usepackage[usenames]{color} 

\usepackage{soul} 

\newcommand{\beq}{\begin{equation}}
\newcommand{\eeq}{\end{equation}}
\newcommand{\bea} {\begin{eqnarray}}
\newcommand{\eea} {\end{eqnarray}}

\begin{document}

\title{Active fluctuations in the harmonic chain: phonons, entropons and velocity correlations}
\author{U. Marini Bettolo Marconi}
\affiliation{INFN Sezione di Perugia, I-06123 Perugia, Italy}

\author{H. L\"owen}
\affiliation{Heinrich Heine Universit\"at D\"usseldorf, D\"usseldorf, Germany. }

\author{L. Caprini}
\affiliation{University of Rome La Sapienza, Rome, Italy. }
\affiliation{Heinrich Heine Universit\"at D\"usseldorf, D\"usseldorf, Germany. }

\begin{abstract}
Non-equilibrium random fluctuations of non-thermal nature are a salient feature of active matter. In this work, we consider the collective excitations of active systems at high density, focusing on a one-dimensional chain of elastically coupled inertial particles, whose activity is modeled with an Ornstein-Uhlenbeck process. Their excitation spectrum shows the presence of two kinds of fluctuations: the first ones are thermally excited phonons analogous to those of a passive crystal while the second ones have been termed entropons because associated with the entropy production due to the active forces. These two types of fluctuations show different properties: in fact, only entropons generate spatial velocity correlations and fail to satisfy a standard fluctuation-response relation. We derive the exact expression for the equal-time velocity and displacement correlations as well as for the structure factor and in each case identify the phonon and entropon contributions. Finally, we investigate the dynamical properties of the excitations in terms of steady-state two-time correlations, such as the intermediate scattering function and the mean-square displacement, and show that both phonon and entropon fluctuations are characterized by a long wavelength overdamped regime and a short wavelength underdamped regime. In the large persistence case, entropons decay slower than phonons, and in general, activity tends to suppress the oscillations typical of the underdamped regime.

\end{abstract}

\maketitle


\section{Introduction}
Active matter is an exciting branch of physics that is highly relevant to biology since it helps to clarify the motion and self-organization of living organisms~\cite{marchetti2013hydrodynamics, elgeti2015physics, bechinger2016active}.
It comprises living systems, such as spermatozoa, cells and tissues, and animals, but also artificial systems, such as active colloids, microrobots, and active granular particles.
These active units differ from equilibrium matter because they self-propel by converting energy from the environment into persistent motion and are permanently out-of-equilibrium.
Specifically, they maintain their velocity orientation for a characteristic, persistence time, beyond which they randomly change direction. Therefore, activity typically induces ballistic behavior at short or intermediate times and a diffusive behavior at long times.
Remarkably, the mechanism providing the energy necessary to sustain their movement acts individually and independently on each of them, in contrast with more conventional non-equilibrium systems displaced from equilibrium globally by an external force or forced at the boundaries, as in the case of shear driving or thermal gradients.
At a mesoscopic level, the persistence of the self-propulsion can be represented by a stochastic force whose value at a given instant is correlated with the values assumed in a previous time interval.
Its explicit form depends on the description adopted and includes the Active Brownian particle (ABP)~\cite{fily2012athermal, caporusso2020motility, caprini2022role, hecht2022active, omar2023mechanical}, the Active Ornstein-Uhlenbeck particle~\cite{szamel2014self, farage2015effective, maggi2015multidimensional, marconi2015towards, flenner2016nonequilibrium, fodor2016far, sharma2017escape, caprini2018activeescape, woillez2020nonlocal} and Run\&Tumble models~\cite{tailleur2008statistical, angelani2014first, sevilla2019stationary, mori2021first}.

As a consequence of the active force memory, the tools~\cite{gardiner1985handbook} developed for the study of Markov processes and employed to describe the colloidal particle motion in equilibrium with the environment, are usually not applicable. Indeed, activity gives rise to important dynamical differences such as the breaking of the Time Reversal Symmetry (TRS)~\cite{o2022time} and the lack of the detailed balance condition~\cite{klein1955principle,gnesotto2018broken}. This has important repercussions, such as the absence of an equation of state~\cite{solon2015pressure} and thermodynamic potentials, but also non-vanishing entropy production~\cite{maes2003time} and lack of a fluctuation-response relation~\cite{kubo1966fluctuation,harada2005equality}. Whereas the physics of a single active particle is pretty well understood, the active many-body aspects are still the object of vivid interest because it is challenging to understand the interplay between memory and interparticle interactions.
Together, these two mechanisms lead to the emergence of characteristic non-equilibrium phenomena such as motility-induced phase separation~\cite{cates2015motility}, flocking~\cite{cavagna2018physics}, and spontaneous velocity alignment~\cite{caprini2020spontaneous} with emergent spatial velocity correlations~\cite{caprini2020hidden}.
Despite the simplicity of the models, analytical solutions are usually not accessible since the systems are far from equilibrium.
The majority of the investigations have been conducted by numerical simulation methods, whereas a minority used analytical approaches, either based on the mean-field approximation of the many body terms or obtained by using models amenable to an exact mathematical treatment.
In particular, the AOUP model is very versatile and lends itself to analytical developments even in the case of interacting active particles~\cite{marconi2016velocity, martin2021statistical}.
Understanding the role of strong correlations in self-propelled systems is one of the most important challenges in active matter.
In this context, one-dimensional systems comprising many active particles with repulsive interactions have proven to be a useful testbed for the study of the behavior of active matter at high density. They offer a simple yet nontrivial model solvable with reasonable theoretical effort and allow for the direct comparison between numerical simulations and analytical results.

In this paper, we consider the dynamics of an active solid comprised of elastically coupled out-of-equilibrium units~\cite{baconnier2022selective, xu2023autonomous}.
The self-propulsions are described by colored noise forces and act together with white noise thermal forces due to the presence of a low-viscosity solvent.
Compared to previous numerical studies~\cite{locatelli2015active, slowman2016jamming, barberis2019phase, gutierrez2021collective}, here we consider a homogeneous system at high density.
The properties of an active solid are obtained by approximating the complex many-body interactions with a harmonic potential in the spirit of the Debye approach used in solid state physics to derive the specific heat due to the crystal vibrations~\cite{ashcroft2022solid}.
To proceed with the smallest amount of approximations, we consider the one-dimensional version of the model and focus on various properties for which we develop analytical predictions:
we obtain the exact expressions of the correlators of the displacement and velocity as a function of frequency and wavevector.
In particular, we discuss the detailed structure of the correlators and relate it to the excitations of the solid.
The distance from the equilibrium is quantified by measuring the entropy production rate (EPR)~\cite{seifert2012stochastic} of the system which in turn is connected with the observed deviation from the fluctuation-response relation.
Compared to previous studies~\cite{caprini2023entropons, caprini2023entropy}, we also derive the expressions of the correlation functions as a function of time and position and the static structure factor and we analytically predict and discuss the intermediate scattering function and the single-particle mean square displacement.

The structure of the paper is the following: in Sec.~\ref{Sec:Model} we present the model of the 1D active chain, while in Sec.~\ref{omegacorrelation} we introduce the spectral representation of the displacement and velocity correlation functions as well as their relation with the entropy production.
In Sec.~\ref{Statics} we derive the equal-time correlation function, the static structure factor, and in Sec.~\ref{timedependent} we illustrate the behavior of the two-time correlation functions.
Finally, in Sec.~\ref{SecConclusions}, we draw the conclusions and future perspectives of our work.
To make the paper more readable but also self-contained, we relegated some mathematical derivations in the appendices.

\section{Inertial solid AOUP Model}
\label{Sec:Model}

We study a one-dimensional active system consisting of $N$ pure-repulsive inertial self-propelled particles in one dimension.
Particles have mass $m$ and are subject to viscous friction and thermal noise, dissipating and injecting energy respectively.
Experimentally, similar periodic structures formed by microparticles immersed in a low-viscosity solvent have been realized using holographic optical tweezers~\cite{yao2009underdamped}.
In this paper, we include an active force, modeled via Ornstein-Uhlenbeck process, acting on each particle
of the chain~\cite{caprini2021inertial, nguyen2021active}.
The following coupled equations involving the positions, velocities, and active forces, $x_n(t)$ and $v_n(t)$, $f^a_n(t)$, respectively, govern the dynamics of the particles
\begin{eqnarray}&&
\label{dynamicequation0}
\dot x_n(t)=v_n(t)\\&&
\label{dynamicequation1}
m\dot v_n(t) = -m\gamma v_n(t) + \sqrt{2 m\gamma T} \xi_n^t(t) +F_n +f^a_n(t)\, .
\end{eqnarray}
The solvent exerts on the particles a drag force proportional to their velocities with drag coefficient $\gamma$ plus
a white noise random impulsive force due to the thermal agitation proportional to $\xi_n^{t}$.
This process has zero average and time correlation $\delta(t-t')\delta_{nn'}$.
For vanishing activity, such that $f^a_n=0$, the resulting dynamics describe an equilibrium system at a temperature $T$ after imposing the standard Fluctuation-Dissipation relation between drag and stochastic noise strength.
On the other hand, to account for the persistence of the active force we assume the following correlation function
\begin{equation}
\langle f_n^a(t) f_{n'}^a(t') \rangle= (m\gamma v_0)^2\delta_{nn'} e^{-|t-t'|/\tau} \, ,
\label{facorrelation}
\end{equation}
and a vanishing average, $\langle f_n^a(t) \rangle=0$.
The parameter $\tau$ represents the persistence time and $v_0$ the active speed.
For later use, we introduce the active temperature, 
\begin{equation}
T_a=m\tau\gamma v_0^2 \,,
\end{equation}
which is related to the effective diffusion coefficient induced by the activity of a single self-propelled particle.
A well-known procedure to obtain $f^a(t)$ consistent with Eq.~\eqref{facorrelation} is to employ the following Ornstein-Uhlenbeck process:
\begin{equation}
\label{dynamicequation2}
\dot f^a_n(t) =-\frac{1}{\tau} f^a_n(t)+ m\gamma v_0 \sqrt{\frac{2}{\tau}} \xi_n^a(t) \, ,
\end{equation}
where $\xi_n^a(t)$ is a second independent white noise with the same characteristics as $\xi_n^t(t)$.
Notice that the non-reciprocal coupling between the position $x_n$ and the active force $f^a_n$, expressed by Eqs.~\eqref{dynamicequation0}-\eqref{dynamicequation1} and~\eqref{facorrelation}, determines the non-equilibrium behavior of the system.

Finally, the motion of any given particle is influenced by the configuration and the motion of the other particles because
the time-independent force $F_n$ describes the interaction between particles such that $F_n= - \partial_{x_n} U_{tot}$, where the total potential $U_{tot}=\sum_{n=1}^{N} U(|x_{n+1} - x_{n}|)$ is given by the sum of pairwise potentials. In recent numerical work, we assumed that $U$ was a truncated and shifted Lennard-Jones (LJ) potential, namely
\begin{equation}
U(r) = 4 \epsilon \left[ \left(\frac{d}{r}\right)^{12}- \left(\frac{d}{r}\right)^6 \right]+\epsilon \,,
\end{equation}
where $\epsilon$ is an energy scale and $d$ is the nominal particle diameter.
At very high density, under the action of the harsh repulsive forces, the particles tend to form a regular one-dimensional lattice whose nodes denoted by $x^0_n$ are separated by a lattice distance $\bar x$.
To proceed analytically, we have considered the following
approximate description of the model~\eqref{dynamicequation0} consisting of replacing the full non-linear potential by
\begin{equation}
U_{tot} \approx m\omega_E^2 \sum_{n}^N (x_{n+1}-x_n-\bar{x})^2 \,,
\label{eq:potential}
\end{equation}
where $\omega_E^2= \frac{U''(\bar{x})}{m} $ is the Einstein frequency of the solid.
The approximation in Eq.~\eqref{eq:potential} assumes that the excursion of each particle from its lattice node position is not too large, and the interaction is limited to first neighbors only.
It is convenient to switch to displacement coordinates and introduce $u_n\equiv (x_n-x^0_n)$ and adopt periodic boundary conditions throughout the paper.
Since the resulting Langevin equation of the harmonic chain~\eqref{eq:potential} is linear and diagonalizable via Fourier analysis we can determine all the one-time and two-time correlation functions.

Using the double Fourier transform in time and space (see Appendix~\ref{app:spectral} for their definition), we decouple the dynamics for the displacement $u_n$ into its normal modes:
\begin{equation}
\label{eq:activedynamicsb}
-\omega^2 \tilde u_q(\omega)= -i\omega \gamma \tilde u_q(\omega) -\omega_q^2 \tilde u_q(\omega) +\sqrt{\frac{2 T \gamma}{m}} \, \tilde \xi^t_q(\omega)+ \frac{\tilde f^a_q(\omega)}{m}
\end{equation}
where the tilted variables are the time-Fourier transforms of the variables featured in Eqs.~\eqref{dynamicequation0}-\eqref{dynamicequation2} and $\tilde{\xi}^t_q(\omega)$ is the Fourier transform of the white noise which satisfies the relation $\langle \tilde{\xi}^t_q(\omega) \tilde{\xi}^t_{q'}(\omega') \rangle = \delta(\omega+\omega')\delta_{qq'}$.
Moreover, we have introduced the frequency $\omega_q$ of the mode $q$ as:
\begin{equation}
\label{eq:def_omega2}
\omega_q^2 =
4 \omega_E^2 \sin^2(q/2) \,.
\end{equation}
We note that, in the limit of $N\gg 1$, the wavevector $q$ becomes a continuous variable in the interval $-\pi$ to $\pi$ and the Fourier transform of the velocity is simply given by $\tilde v_q(\omega)=i\omega \tilde u_q(\omega)$.
With the help of the response function
\begin{equation}
\mathcal{R}_{\hat{u}\hat{u}}( q,\omega) =-\frac{ 1}{-\omega^2+\omega_q^2 -i\omega\gamma } \,,
\label{responsefunction}
\end{equation}
Eq.~\eqref{eq:activedynamicsb} can be rewritten in a more convenient way as
\begin{equation}
\tilde u_q(\omega)= \mathcal{R}_{\hat{u}\hat{u}}( q,\omega)
\left(\sqrt{\frac{2 T \gamma}{m}} \, \tilde \xi^t_q(\omega)+ \frac{\tilde f^a_q(\omega)}{m}\right) \,,
\end{equation}
which is particularly suitable to calculate dynamical correlation functions in Fourier space.

\subsection{Entropy production of active particles}

The degree of irreversibility of the process induced by the active forces and its distance from thermodynamic equilibrium is measured by the entropy production rate~\cite{tietz2006measurement, seifert2012stochastic, dabelow2019irreversibility, fodor2022irreversibility, o2022time}.
The breaking of the time-reversal symmetry in off-equilibrium systems implies that a particular sequence of mesoscopic states, as those described by our Langevin equations, has a different
probability with respect to the time-reversed sequence.
This asymmetry is expressed through the Kullback-Leibler divergence, i.e. the logarithm of the ratio between the probability weight, $P_f$, associated with a path forward in time and the weight $P_r$ for the time-reversed path:
\begin{equation}
\dot{S} = \lim_{t\to\infty}\frac{1}{t}\left\langle\ln{\left(\frac{P_f}{P_r}\right)}\right\rangle \, .
\label{eq:EPR}
\end{equation}
The time derivative of the Kullback-Leibler divergence defines the entropy production rate~\cite{seifert2012stochastic, spinney2012entropy, fodor2016far, caprini2019entropy, herpich2020stochastic, ferretti2022signatures, suchanek2023entropy}.
With such a tool one easily discriminates between non-equilibrium steady states (NESS) and equilibrium states, where $\dot S=0$. Using the formalism of stochastic thermodynamics \cite{seifert2012stochastic, speck2016stochastic, forastiere2022linear},
as shown in appendix~\ref{EPRapendix}, the entropy production for the AOUP is proportional to the power injected by the active forces, $f_n^a$, on the particles, and vanishes if the persistence time goes to zero and the dynamics become reversible~\cite{sekimoto2010stochastic}.
Hence, the EPR for the AOUP turns out to be the ratio between the power dissipated from the particle to the environment under the form of heat and the temperature, $T$, of the surroundings and reads~\cite{shankar2018hidden, dabelow2021irreversible, caprini2023entropy}
\begin{equation}
\label{eq:entropyprodrate}
\dot S=\frac{1}{T}\sum_n\langle f_n^a(t) v_n(t) \rangle \,.
\end{equation}
In the next section, we study the fluctuations of the active system and with the help of the EPR separate their equilibrium from the non-equilibrium contributions. To achieve this goal, we consider the correlation functions, the response function, and the spectral entropy production.


\section{Dynamical correlations and Entropy production}
\label{omegacorrelation}
To compute the correlation functions we consider the average
of the bilinear combinations
of the displacement and velocity variables. We define the following displacement-displacement correlation
in the $(q,\omega)$ representation:
\begin{equation}
2\pi \delta(\omega+\omega')\Phi_{uu}^{total}(q,\omega)=\langle  \tilde u_q(\omega)\tilde u_{-q}(\omega') \rangle \,, 
\label{eq:phiuuom}
\end{equation}
where the angular brackets symbolize the averages over the realizations of both the thermal noise and active forces.
The corresponding time-correlation in the $(q,t)$ representation is obtained from Eq.\eqref{eq:phiuuom} through the Fourier transform:
\begin{equation}
C_{uu}^{total}(q,t)=\int_{-\infty}^\infty \frac{d\omega}{2\pi}\Phi_{uu}^{total}(q,\omega)\, e^{-i\omega t} \,.
\label{eq:phiuuot}
\end{equation}
The following Fourier spatial transform gives the spatio-temporal correlation:
\begin{equation}
c_{uu}^{total}(n,t)=\int_{-\pi}^\pi \frac{dq}{2\pi}\, C_{uu}^{total}(q,t)\, e^{-i q n} \,.
\label{eq:phiuuotspace}
\end{equation}
In the rest of the paper,
the remaining correlation functions involving the velocity and the partial contributions to the total quantities satisfy relations analogous to Eqs.~\eqref{eq:phiuuot}-\eqref{eq:phiuuotspace}.

By using the linearity of Eq.~\eqref{eq:activedynamicsb}, we split the displacement-displacement correlation function into the sum of two parts, the phonon contribution and the entropon~\cite{caprini2023entropons, caprini2023entropy} contribution, corresponding to the modes excited by the thermal noise and the active force, respectively:
\begin{equation}
\Phi_{uu}^{total}(q,\omega)=\Phi_{uu}^{phonon}(q,\omega)+\Phi_{uu}^{entropon}(q,\omega) \,,
\label{eq:decomposition}
\end{equation}
where
\begin{eqnarray}&&
\Phi_{uu}^{phonon}(q,\omega)=
\frac{T}{m}\frac{ \gamma}{(\omega^2-\omega_q^2 )^2 + \omega^2\gamma^2 }
\label{eq:correlationphonon}
\\&&
\Phi_{uu}^{entropon}(q,\omega)=
\frac{T_a}{m} \frac{ 1 }{ 1+ \omega^2\tau^2 }\frac{ \gamma}{(\omega^2-\omega_q^2 )^2 + \omega^2\gamma^2 }\,.
\label{eq:correlationentropon}
\end{eqnarray}
We have identified the first term with the contribution due to the (underdamped) phonons and the second with the one due to the entropons, i.e.\ the excitations associated with the entropy production of the system.
As discussed in our previous work~\cite{caprini2023entropons, caprini2023entropy}, at the origin of such terminology is the so-called Harada-Sasa relation~\cite{harada2005equality} involving three subjects: the displacement fluctuations $ \Phi_{uu}(q,\omega)$, the imaginary part of the response function, $\mathcal{R}_{\hat{u}\hat{u}}$ and spectral entropy production $\sigma(q,\omega)$.
\begin{equation}
\frac{ \Phi_{uu}^{total}(q,\omega) }{T} =-
\frac{ \text{Im} [\mathcal{R}_{\hat{u}\hat{u}}(q, \omega)]}
{\omega}
+ \frac{\sigma(q,\omega)}{m \omega^2\gamma}\, .
\label{eq:haradasasa}
\end{equation}
According to Eq.~\eqref{eq:haradasasa}, in active systems, the displacement correlation does not satisfy the usual fluctuation-response relation between fluctuation and response but is subject to an extended relation involving a third observable, the spectral entropy production, as also discussed in the framework of active field theories~\cite{nardini2017entropy}. To derive 
Eq.~\eqref{eq:haradasasa} from Eq.~\eqref{eq:decomposition} we remark the following equality
\begin{equation}
\dot S=
\frac{\sum_n \langle v_n(t) f^a_n(t)\rangle}{T}= \frac{m\gamma}{T} \sum_q \int_{-\infty}^\infty\, \frac{d\omega}{2\pi} \, \omega^2\, \Phi_{uu}^{entropon}(q,\omega) \,,
\label{power75b}
\end{equation}
and consider the spectral representation of the EPR, i.e. its decomposition in independent Fourier modes
\begin{equation}
\dot S=
\sum_q \int_{-\infty}^\infty\, \frac{d\omega}{2\pi} \sigma(q,\omega) \,.
\label{eprspectral}
\end{equation}
By comparing Eqs. \eqref{power75b} and \eqref{eprspectral}, we obtain the equality
\begin{equation}
\sigma(q,\omega)=m \gamma \omega^2 \, \frac{ \tilde \Phi_{uu}^{entropon}(q,\omega)}{T} \,.
\end{equation}
Considering the response function~\eqref{responsefunction} we derive a second relation
\begin{eqnarray}&&
\frac{ \Phi_{uu}^{phonon}(q,\omega)}{T}=
-\frac{ \text{Im} [\mathcal{R}_{\hat{u}\hat{u}}(q, \omega)]}
{\omega}
\end{eqnarray}
and collecting these results we explicitly obtain the Harada-Sasa relation~\eqref{eq:haradasasa}.
We remark that the spectral entropy production, $\sigma(q,\omega)$, provides detailed information on how the various modes dissipate energy. In particular, the shorter the wavelength the smaller the entropy production.
For completeness and because of the applications in the rest of the paper, we establish a relation between the velocity and the displacement correlation functions, $\Phi_{vv}^{total}(q,\omega)=\omega^2\Phi_{uu}^{total}(q,\omega)$, and obtain the following result:
\begin{equation}
\Phi_{uu}^{entropon}(q,\omega)
=\frac{T_a}{T}\Phi_{uu}^{phonon}(q,\omega)-\tau^2\,\Phi_{vv}^{entropon}(q,\omega)\,,
\label{eq:correlationb}
\end{equation}
where
\begin{eqnarray}&&
\Phi_{vv}^{entropon}(q,\omega)=
\frac{T_a}{m} \frac{ \omega^2 }{ 1+ \omega^2\tau^2 }\frac{ \gamma}{(\omega^2-\omega_q^2 )^2 + \omega^2\gamma^2 }
=\frac{T}{m \gamma}\sigma(q,\omega) \,.
\label{entroponomega}
\end{eqnarray}
Such a formula links the presence of dynamical velocity correlations to the irreversibility quantified by the spectral entropy production, in contrast with the behavior of passive systems (corresponding to $T_a=0$) where both the velocity correlations and the EPR vanish.

\begin{figure}[t]
	\includegraphics[width=\textwidth]{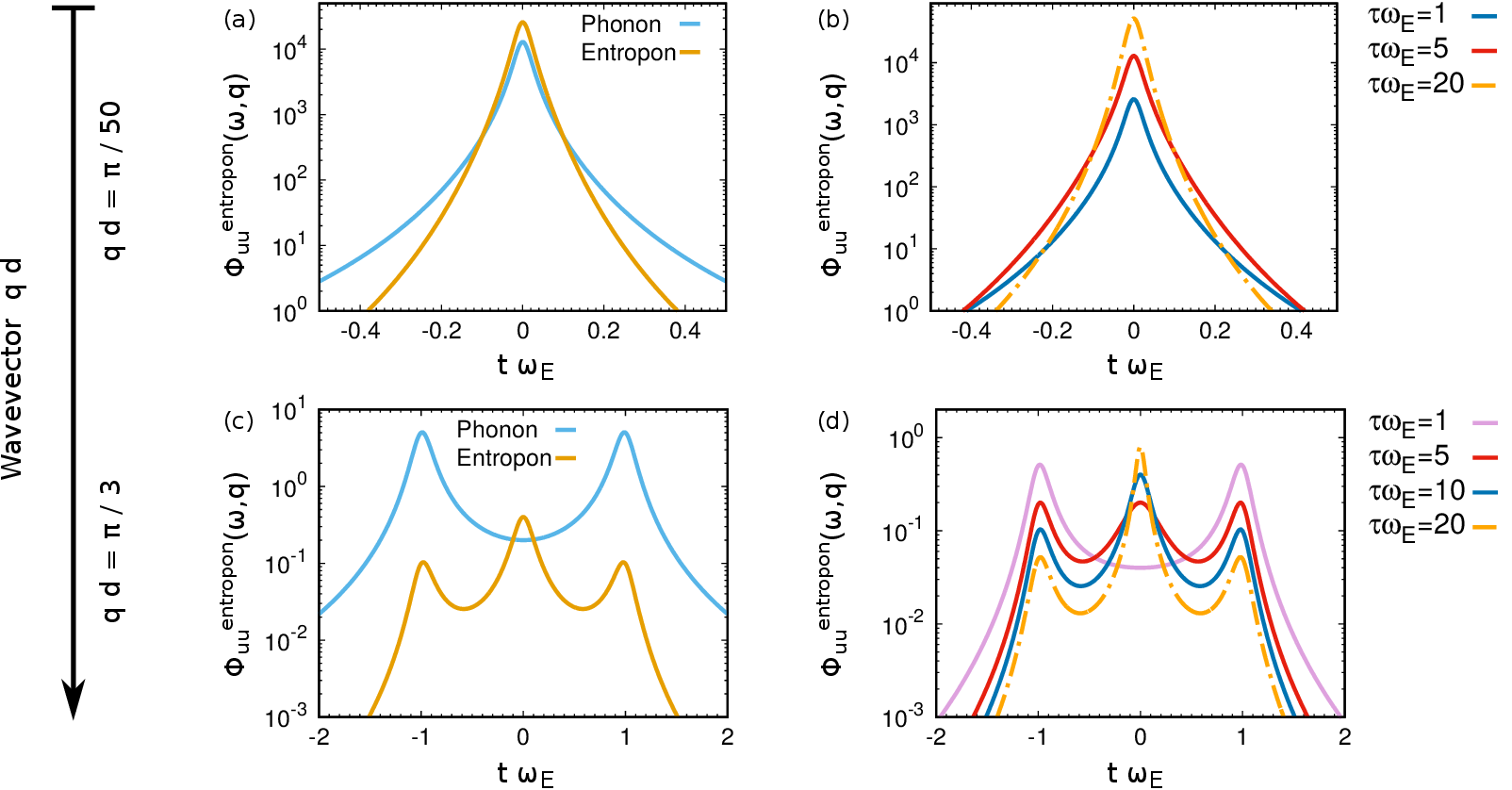}
	\caption{\textbf{Displacement dynamical correlations}, $\Phi_{uu}(\omega,q)$, as a function of the frequency $\omega$ rescaled by Einstein's frequency $\omega_E$.
(a), (b): $S(\omega,q)$ for a subcritical $q$-value, $q\sigma=\pi/50$.
(c), (d): $S(\omega,q)$ for a supercritical $q$-value, $q\sigma=\pi/3$.
In (a) and (c), the phonon contribution to $\Phi_{uu}(\omega,q)$ is compared with the entropon one for $\tau\omega_E=10$. In (b) and (d), $\Phi_{uu}^{entropon}(\omega,q)$, i.e. the entropon contributions to $\Phi_{uu}(\omega,q)$, are shown for different values of the reduced persistence time $\tau\omega_E$, as reported in the external legends.
Colored curves are obtained by plotting Eq.\eqref{eq:correlationphonon} (phonons) and Eq.\eqref{eq:correlationentropon} (entropons).
The other parameters are: $m/\gamma \omega_E=5$, $T / (\omega_E^2 d^2)=0.1$, and $v_0 /(\omega_E d)=1$.
	}
	\label{fig:fig1}
\end{figure}

Returning to Eq.~\eqref{eq:decomposition}, we immediately see that the character of the eigenmodes of the system strongly depends on the value of the wavevector $q$. 
Modes with larger values of $\omega_q$ are more likely to be underdamped.
The crossover from overdamping to underdamping is governed by the ratio of inertial time $1/\gamma$ and relaxation time of the $q$-mode, $1/\omega_q$. It moves towards underdamping by decreasing the drag coefficient or increasing the strength of the potential.
 In fact, the poles of the correlator $\Phi^{phonon}_{uu}(q,\omega)$ in Eq.~\eqref{eq:correlationphonon} lie on the imaginary axis when $\omega_q^2<\gamma^2/4$. Instead, for $\omega_q^2>\gamma/4$ the poles become complex developing a real part  and producing oscillations in the time-dependent correlation functions as we shall illustrate in Section \ref{timedependent}.

The power spectrum of phonon fluctuations changes from an overdamped Lorentzian spectrum with an unimodal shape to an underdamped bimodal shape with the appearance of two symmetric resonance peaks.
The phonon correlation, $\Phi^{phonon}_{uu}(q,\omega)$ displays a single peak centered at $\omega=0$ if $ \omega_q^2$ is less than the critical value $\gamma^2/2$, as shown in In Fig.\ref{fig:fig1} (a) (cyan curve) for $q=\pi/50$.
By contrast, a double peak structure appears for $\omega_q^2> \gamma^2/2$ as shown in Fig.\ref{fig:fig1} (c) (cyan curve) for $q=\pi/3$.
The side peaks occur at frequencies $\omega^*=\pm\sqrt{\omega_q^2-\frac{\gamma^2}{2}}$ and are the remnant of the phonon peaks at frequencies $\pm \omega_q$ of the corresponding frictionless system ($\gamma=0$).
We name them Brillouin peaks, bearing in mind that only modes having wavevectors larger than $q_{crit}=2\arcsin(\frac{\gamma}{2^{3/2}\omega_E}) $ present this type of structure in the spectrum.

We, now, shift our focus to the second contribution to the displacement fluctuation, the entropon, $\Phi_{uu}^{entropon}(q,\omega)$: the presence of the extra factor $(1+ \omega^2\tau^2)^{-1}$, due to the active force correlation, depletes the peaks at $\omega^*=\pm\sqrt{\omega_q^2-\frac{\gamma^2}{2}}$ but produces an extra peak at $\omega=0$ in the total displacement correlation function.
However, when $\omega_q^2<\gamma^2/2$ such an extra contribution occurs in the same region $\omega\approx 0$ where also the phonon central peak is located, as shown in Fig.\ref{fig:fig1} (a) (yellow curve). Therefore, it becomes visually appreciable only above the critical value $\omega_q^2>\gamma^2/2$ as revealed in as shown in Fig.\ref{fig:fig1} (c) (yellow curve).
In conclusion, in the case of phonons, the harder modes, i.e. those having a wavevector $q$ above the critical value, $q_{crit}$, display a Brillouin double peak structure as a function of $\omega$.
Instead, in the same range of parameters the supercritical entropons show a triple-peak symmetric structure.
As we shall show in Sec.~\ref{timedependent}, the multipeak structure in the frequency representation of the correlation functions reflects the presence of different time scales in the relaxation of the fluctuations.

From Eq.~\eqref{eq:correlationentropon}, it is evident that the swim velocity $v_0$ trivially affects $\Phi_{uu}^{entropon}(q,\omega)$ by simply increasing the amplitude of this correlation through the prefactor $T_a$, identified as the active temperature.
The main changes due to activity occur by increasing the persistence time, $\tau$ normalized with the Einstein frequency $\omega_E$.
For overdamped modes, such that $\omega_q^2<\gamma^2/2$, the persistence increase induces a higher and narrow central peak as shown in Fig.~\ref{fig:fig1}~(b).
For underdamped modes, such that $\omega_q^2>\gamma^2/2$, the decrease of $\tau$ reduces the height of the additional central peak, as revealed in Fig.~\ref{fig:fig1}~(d). This reduction continues until this peak is suppressed in the small persistence regime when active particles are effectively passive and $\Phi_{uu}^{entropon}(q,\omega)$ is only characterized by the two lateral peaks.

 \section{Static correlations}
 \label{Statics}

  We, now, consider the steady state of the system and determine its stationary properties including the equal time values of the displacement and velocity correlations and the static structure factor.
  To this purpose, we integrate over the frequency the corresponding $\omega$ correlations according to the prescription:
  \begin{equation}
  C_{xx}^{total}(q,t)=\int_{-\infty}^\infty \frac{d\omega}{2\pi} \, e^{-i\omega t}\, \Phi_{xx}^{total}(q,\omega) 
  \label{timecorrelations}
  \end{equation}
   where the subscript  $\{xx\}$ stands for $\{vv\}$ or $\{uu\}$.
 
\subsection{Spatial velocity correlations}

\begin{figure}[t]
	\includegraphics[width=0.85\textwidth]{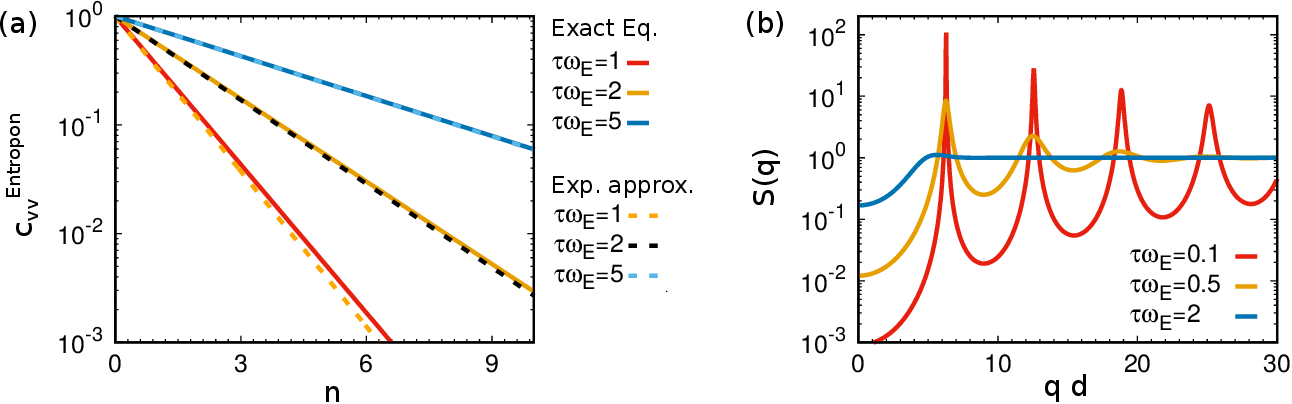}
	\caption{\textbf{Static correlations.} (a): Spatial velocity correlation, $c_{vv}^{entropon}$, due to entropons as a function of the dimensionless distance $n$. $c_{vv}^{entropon}$ is normalized at one and is shown for different values of the persistence time $\tau\omega_E$ normalized by Einstein's relation.
Solid lines are obtained by plotting the exact expression~\eqref{eq:spatial_corr_theory2}, while dashed lines are the results of the exponential approximation~\eqref{eq:spatial_corr_approx}.
(b): Static structure factor, $S(q)$, as a function of wavevector $q$ normalized with the particle diameter $d$, for several values of $\tau\omega_E$.
The other parameters are: $m/\gamma \omega_E=5$, $T / (\omega_E^2 d^2)=0.1$, and $v_0 /(\omega_E d)=1$. 	}
	\label{fig:vel_corr}
\end{figure}

Let us begin with the steady-state velocity correlation function, i.e. corresponding to $t=0$.
One of the most striking aspects of active systems is the presence of correlations between the velocities of different particles which are completely absent at equilibrium~\cite{caprini2020spontaneous, caprini2020hidden}.
 Notwithstanding the absence of any alignment interaction, the velocities of different particles  become spatially correlated over distances up to the correlation length, $\lambda$, in other words, they form domains where the velocities display a certain degree of  coherence. This phenomenon occurs at high density, in phase-separated~\cite{caprini2020spontaneous, yang2023coherent}, solid~\cite{caprini2021spatial, abbaspour2023long} and liquid configurations~\cite{caprini2020hidden, szamel2021long, marconi2021hydrodynamics, debets2023microscopic} as well as in active glasses~\cite{szamel2015glassy, henkes2020dense, keta2022disordered, debets2023glassy, keta2023emerging} and systems governed by feedback mechanisms~\cite{kopp2023spontaneous}.
In the steady regime, the equal-time velocity correlation function of each mode $q$ is:
 \begin{eqnarray}
 C_{vv}^{total}(q,0)=
  \frac{T}{m}
 + \frac{T_a}{m}  \Bigl[
 \frac{1}{1 +\tau\gamma + \tau^2 \omega_q^2 }
  \Bigr] \,.
  \label{eq:velocitycorrel}
  \end{eqnarray}
Considering~\eqref{eq:def_omega2} and using the small $q$ expansion of formula~\eqref{eq:velocitycorrel}
we obtain an Ornstein-Zernike expression for the equal time  $q$-correlation characterized by
 the non dimensional correlation length, $\lambda$,  (because expressed in lattice units)
 from the relation
\begin{equation}
\lambda^2= \frac{\omega_E^2\tau^2}{1+\tau\gamma }\, .
\end{equation}
As already noticed~\cite{caprini2021spatial} such a length is independent of $T_a$, the intensity of the active force, whereas
it is an increasing function of the Einstein frequency $\omega_E$, the persistence time $\tau$ and a decreasing function of the damping time $1/\gamma$.
To characterize the size of the velocity domains, we consider the equal-time  spatial velocity correlation functions, $\langle v_n(0) v_0(0) \rangle$, following a strategy similar to Ref.~\cite{caprini2020hidden}.
We go back to the real space description by integrating the equal-time velocity q-correlation with respect to $q$:
\begin{equation}
c^{total}_{vv}(n,0) =\int_{-\pi}^\pi  \frac{dq}{2\pi} e^{-iqn}\, C_{vv}^{total}(q,0)\, .
\end{equation}
If subject to pure thermal noise,  the velocities of different particles (i.e. $n\neq 0$) are not correlated and we find:
 \begin{equation}
 \frac{T}{m}\int_{-\pi}^\pi  \frac{dq}{2\pi} e^{-iqn} =\frac{T}{m}\delta_{n,0}\,.
\end{equation}
  On the other hand, the entropon component, $\Phi_{vv}^{entropon}(q,\omega)$, proportional to the intensity of the active noise, $T_a$, gives rise to spatial velocity correlations. For zero separation $n=0$, 
the equal-time space velocity-velocity correlation reads:
\begin{equation}
c^{total}_{vv}(0,0) =\frac{T}{m}+\frac{\frac{T_a}{m} }{1+\tau\gamma}  \frac{1}{\left(1+4\lambda^2  \right)^{1/2}} \,.
\label{cvvtempiuguali}
\end{equation}
It is worth noticing that
the total entropy production rate $\dot S$ is proportional to the
amplitude of the velocity correlations, $c^{entropon}_{vv}(0,0)$, in virtue of Eq.~\eqref{eprspectral} and~\eqref{entroponomega}.
If compared with the EPR of a single free particle, $\dot S_1=\frac{\frac{T_a}{m} }{1+\tau\gamma} $ the EPR per particle of the one dimensional harmonic solid is smaller by a factor
$(1+4\lambda^2  )^{-1/2}$, i.e. the Einstein frequency reduces the EPR.    
For generic values of the separation $n$ it is possible to obtain the exact correlation function
(see appendix~\ref{spatialintegrals} for details):
\begin{equation}
\label{eq:spatial_corr_theory2}
c^{entropon}_{vv}(n,0)=         c^{entropon}_{vv}(0,0)\Bigl[\frac{1+2\lambda^2-\sqrt{1+4\lambda^2}}        {2\lambda^2}\Bigr]^n
\end{equation} 
In the limit of $\lambda\gg1$, expression~\eqref{eq:spatial_corr_theory2} can be approximated as
\begin{equation}
\label{eq:spatial_corr_approx}
c^{entropon}_{vv}(n,0)\approx  c^{entropon}_{vv}(0,0) \exp{\left(-\frac{n}{\lambda}\right)}
\end{equation}
where the typical length, $\lambda$, associated with the exponential decay of the velocity correlation,  represents the size of the velocity domains. 
Such a an average size of  scales as $\sim \tau^{1/2}$ and increases when the density increases, because $\omega_E$
increases as the lattice spacing, $\bar{x}$, decreases. 
The theoretical expression for the spatial velocity correlations, Eq. \eqref{eq:spatial_corr_theory2}, is reported in Fig.~\ref{fig:vel_corr}~(a) together with the exponential approximation, Eq. \eqref{eq:spatial_corr_approx}.
In general, the two expressions are in good agreement as the persistence time (and thus $\lambda$) increases.

 \subsection{Equal-time displacement  correlation} 
Following an integration procedure similar to the one adopted in the analysis of the velocity correlations we obtain the static displacement $q$-correlation
  \begin{eqnarray}&&
  C_{uu}^{phonon}(q,0)
  = \frac{T}{m}\frac{1}{\omega_q^2} \,,
  \\&&
  C_{uu}^{entropon}(q,0)
  =
   \frac{T_a}{m}\frac{1}{\omega_q^2}  \frac{1+\tau\gamma}{  1+\tau\gamma+\omega_q^2\tau^2    }\,.
  \end{eqnarray}
  We express the steady-state displacement fluctuations of a tagged particle as:
\begin{equation}
 \langle u_n^2\rangle =\int_{2\pi/N}^{\pi}  \frac{dq}{\pi} \,
C_{uu}^{tot}(q,0)=\int_{2\pi/N}^{\pi}  \frac{dq}{\pi} \,\Bigl(\frac{(T+T_a)}{m} \frac{1}{\omega_q^2}-\frac{T_a}{m}\frac{\tau^2}{1+\tau\gamma+\tau^2\omega_q^2}\Bigr) \,.
\label{cuuintegral}
\end{equation}
Since $\omega_q^2$ behaves as $q^2$ for small $q$ values the integral diverges in the $N\to \infty$ limit, 
we introduce the lower limit $2\pi/N$  to discuss  the dependence of the fluctuations on the size $N$ of the system. 
The integration yields the following result:
\begin{equation}
\langle u_n^2\rangle=\frac{N} {2\pi^2}\frac{(T+ T_a)}{m  \omega_E^2} -\tau^2 \frac{T_a}{m}\, 
\frac{1}{1+\tau\gamma}  \frac{1}{\left(1+4\lambda^2 \right)^{1/2}} \,.
\label{notpossible}
\end{equation}
The presence in Eq.~\eqref{notpossible} of the last term, stemming from the velocity correlations,
shows that is not possible to describe the active solid as a hotter crystal, i.e. in terms of a renormalised temperature.

It is useful to express the mean square displacement in terms of the MSD, $a^2$,
of a single particle in equilibrium with a heat bath
at temperature $T+T_a$ and
in a confining harmonic potential having the same Einstein frequency, $\omega_E^2 x^2/2$, of the solid.
The resulting displacement fluctuation diverges linearly with the size $N$ of the chain, with a finite negative correction 
associated with the presence of static velocity correlations:
\begin{equation}
\langle u_n^2\rangle=\frac{N} {2\pi^2}\frac{(T+ T_a) }{m \omega_E^2} -\tau^2 c^{entropon}_{vv}(0,0) \,.
\label{eq:unaverage}
\end{equation}
%
%
The difference between the displacement of particles separated by $n$ lattice
steps reads:
\begin{equation}
\langle(u_{n+m}-u_m)^2\rangle = 2\int_{0}^\pi \frac{dq}{\pi}
\Bigl( 1 - \cos(n q) \Bigr) C^{tot}_{uu}(q,0)\,.
\label{eq:differenceu}
\end{equation}
Using the discrete translational invariance of the lattice we find:
\begin{eqnarray}&&
\langle (u_n-u_0)^2 \rangle=\frac{(T+ T_a) }{m \omega_E^2}
\int_{0}^\pi \frac{dq}{\pi} \, \frac{ 1 - \cos(n q) }{1-\cos(q)}
-2\tau^2 \frac{T_a}{m}\int_{0}^\pi \frac{dq}{\pi} \, \frac{ 1 - \cos(n q) }{1+\tau\gamma+2\tau^2\omega_E^2(1-\cos(q))}
\label{twointegrals}
\end{eqnarray}
and for large values of $n$ obtain the result:
\begin{eqnarray}&&
\langle (u_n-u_0)^2 \rangle
\approx n \frac{(T+ T_a) }{m \omega_E^2} -2\tau^2 \frac{T_a}{m}\,
\frac{1}{1+\tau\gamma} \frac{1}{\left(1+4\lambda^2 \right)^{1/2}}\left(1-\exp{\left(-\frac{n}{\lambda}\right)}\right)\,.
\label{resultintegrals}
\end{eqnarray}
Although $u_n$ and $u_0$ each have infinite fluctuations in the $N\to \infty$ limit, their difference has only finite fluctuations roughly proportional to their separation showing that the local deformations of the lattice are finite. However,
when the separation increases the fluctuation~\eqref{resultintegrals} behaves like a constant times $n$ plus an exponential correction stemming from the velocity correlations.

\subsection{Static structure factor}

We now study the system in terms of one of its collective variables, the density defined as $\hat n(x)=\sum_n \delta(x-x_n)$. The average over thermal noise and active force of its Fourier transform
is easily obtained by remarking that the distribution of the displacement is Gaussian so that the following average can be easily performed:
\begin{equation}
\langle \hat n_q\rangle = \sum_n e^{iqn} \langle e^{iq u_n}\rangle =\sum_n e^{iqn} e^{-q^2\langle u_n^2\rangle}=N \delta_{q,0} \,.
\end{equation}
The last equality states that only the Fourier component with $q= 0$ has a non-vanishing amplitude because according to Eq.~\eqref{eq:unaverage} the amplitude $\langle u_n^2\rangle$ increases linearly with $N$.
Thus the non-equilibrium steady density is uniform and in the limit $N\to \infty$ the system is liquid-like.
However, the static density correlations still show some structure and we use
the equal-time structure factor $S(q)$ to analyze the
existence of order in the chain. $S(q)$ is proportional to the scattering cross-section of an incoming particle with pre-collisional wavevector $k_i$ and final wavevector $k_f$, such that $q=k_f-k_i$
and is related to the pair correlation $g(x)$ by the relation:
\begin{equation}
S(q)=1+\rho\int dx \,(g(x)-1) \cos(qx)
\end{equation}
where $\rho$ is the average number density.
In the Born approximation, $S(q)$ is written as:
\begin{equation}
S(q)=\frac{1}{N} \sum_{l,j=1}^N\exp(-i q(l-j))\left\langle \exp(-i q(u_l-u_j))\right\rangle \,.
\end{equation}
Since the displacement has a Gaussian distribution, by using Eq.~\eqref{resultintegrals} and neglecting
the second subleading term due to velocity correlations, we have
\begin{equation}
S(q)=\frac{1}{N} \sum_{l,j=1}^N\exp\left(-i q(l-j)\right) \exp\left(- q^2 |l-j|\frac{(T+ T_a) }{2 m \omega_E^2} \right) \,,
\end{equation}
where the double sum runs over the indices of the particles.
The double sum can be reduced to a single sum and performed with the following result~\cite{emery1978one}:
\begin{equation}
S(q)=1+2 \sum_{n=1}^{N-1} \left( 1-\frac{n}{N}\right) \cos\left(q n\right) \exp\left(- q^2 n \frac{(T+ T_a) }{2m \omega_E^2} \right)=\frac{ \sinh\left(q^2\frac{(T+ T_a) }{2m \omega_E^2}\right)}{\cosh\left(q^2 \frac{(T+ T_a) }{2m \omega_E^2}\right) -\cos(q)} \,.
\label{eq:strucfactor}
\end{equation}
The structure factor $S(q)$ displays local maxima at positions $q_G=2 k\pi$, where $k$ is an integer number.
The shape of $S(q)$ describes a structured liquid given the absence of long-range translational order in one-dimensional systems. 
Their amplitudes decrease as $(\pi k)^{-2}\frac{m \omega_E^2}{(T+ T_a) }$ as the index $k$ increases as shown in Fig.~\ref{fig:vel_corr}~(b).
Near the Bragg wavevectors $q_G= 2k \pi$ with nonzero $k$, the peaks are nearly Lorentzian of width
$2\pi k^2 \frac{(T+ T_a) }{m \omega_E^2}$, thus the larger the intensity of the active force and the higher the temperature the less pronounced their heights.
Notice that the forward scattering peak at $q=0$ is absent since Eq.~\eqref{eq:strucfactor} gives $S(q\to 0)=\frac{(T+ T_a) }{m \omega_E^2}$ because to derive formula~\eqref{eq:strucfactor} we performed the limit $N\to \infty$. 
However, the limits $N\to\infty$ and $q\to 0$ do not commute~\cite{schonhammer2014quantum} and the correct limit $S(0)=N$ cannot be recovered.

\section{Time-dependent fluctuations}
 \label{timedependent}

\begin{figure}[t]
\includegraphics[width=0.85\textwidth,angle=0]{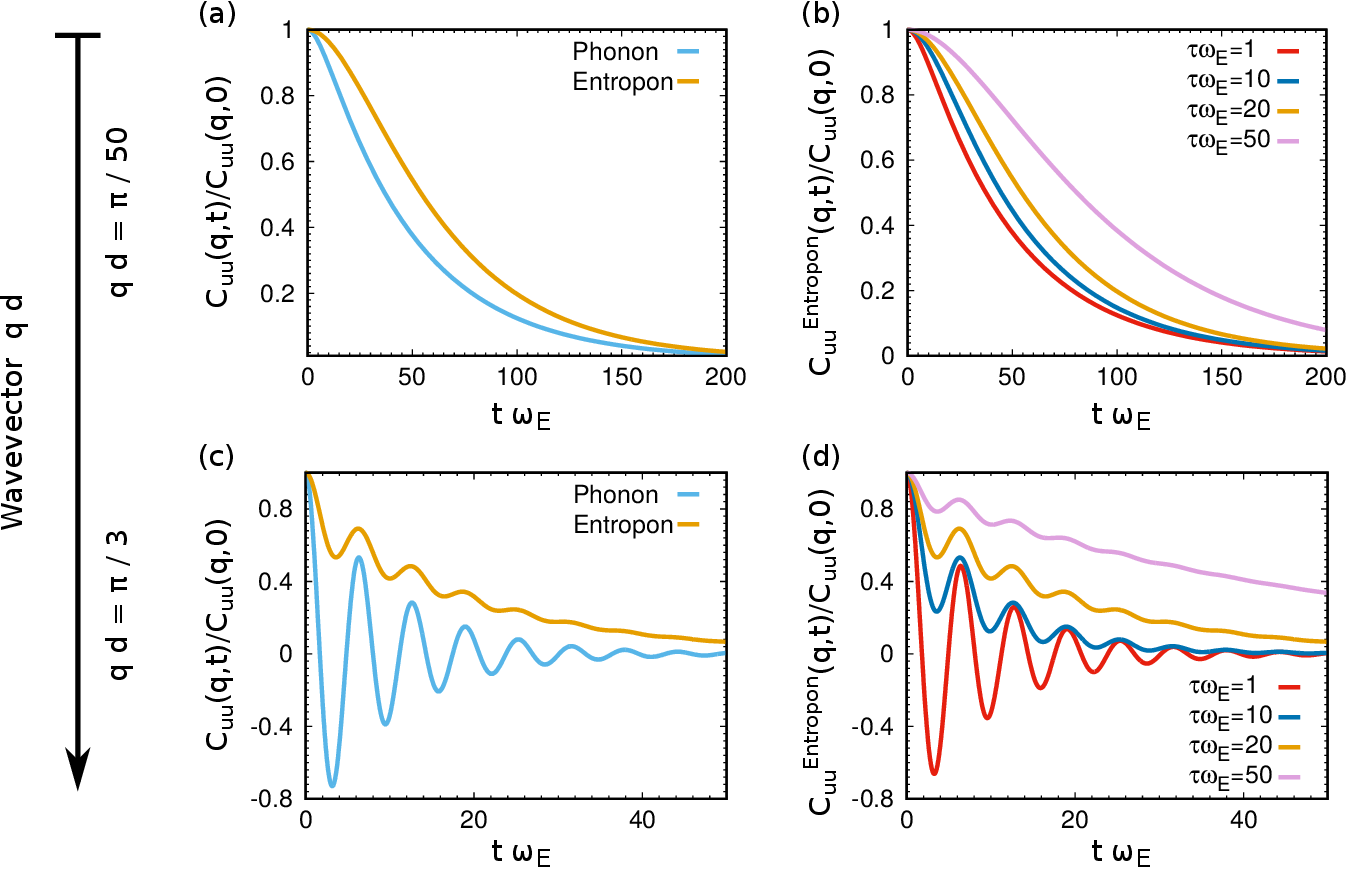}
\caption{{\textbf{Intermediate scattering function}}, $C_{uu}(q,t)$, as a function of time $t$ rescaled by the Einstein's frequency $\omega_E$. Each $C_{uu}(q,t)$ is normalized by its initial time value $C_{uu}(q,0)$.
(a), (b): $C_{uu}(q,t)$ for a subcritical $q$-value, $q\sigma=\pi/50$. 
	(c), (d): $C_{uu}(q,t)$ for a supercritical $q$-value, $q\sigma=\pi/3$. 
	In (a) and (c), the phonon contribution to $C_{uu}(q,t)$ is compared with the entropon one for $\tau\omega_E=10$. In (b) and (d), $C_{uu}^{entropon}(q,t)$, i.e.\ the entropon contributions to $C_{uu}(q,t)$ are shown for different values of the reduced persistence time $\tau\omega_E$, as reported in the external legends.
	Colored curves are obtained by plotting Eq.~\eqref{eq:correlationphonon} (phonons) and Eq.~\eqref{eq:correlationentropon}(entropons).
	The other parameters are:  $m/\gamma \omega_E=5$, $T / (\omega_E^2 d^2)=0.1$, and $v_0 /(\omega_E d)=1$. 
}
\label{fig:Intermediate}
\end{figure}

Spontaneous fluctuations or perturbations due to external agents almost always take place and therefore it is important to know how dense systems of particles relax toward a steady state.
The relevant information is contained in the two-time correlation functions which can be determined by Fourier transforming their spectra obtained in Sec~\ref{omegacorrelation}.
In this way, it is possible to calculate the intermediate scattering function determining the displacement-displacement relaxation of the $q$-mode. In addition, we investigate the single-particle mean-square displacement.

\subsection{Intermediate scattering function}
For $t\geq 0$, by using the Fourier transform~\eqref{timecorrelations}, we find that the total velocity correlator as a function of the wavevector $q$ and time $t$ is the sum of the phonon and entropon contributions:
\begin{eqnarray}&&
\label{eq:phonon}
C_{vv}^{phonon}(q,t)
=\frac{T}{m}
e^{-\gamma t/2}
\Bigl[ {\cal C}(q,t)
- {\cal S}(q,t)\Bigr]
\\&&
C_{vv}^{entropon}(q,t)
=
\frac{T_a}{m} \frac{1}{ (1+\omega_q^2\tau^2)^2-\gamma^2 \tau^2 }
\Bigl\{
e^{-\gamma t/2}
\Bigl[ (1+\omega_q^2\tau^2)\, {\cal C}(q,t)
-(1-\omega_q^2\tau^2)\, {\cal S}(q,t)\Bigr] -\gamma \tau e^{- t/\tau}
\Bigl\}
\label{eq:entropon}
\end{eqnarray}
where we have introduced the abbreviations:
\begin{eqnarray}
&&
{\cal C}(q,t) \equiv \theta(\frac{\gamma^2}{4} -\omega_q^2)\, \cosh(t\sqrt{\gamma^2/4- \omega_q^2}\, )
+ \theta(\omega_q^2-\frac{\gamma^2}{4})\, \cos(t\sqrt{\omega_q^2-\gamma^2/4} )
\\&&
{\cal S}(q,t)\equiv \theta(\frac{\gamma^2}{4} -\omega_q^2) \, \frac{\frac{\gamma}{2}}{\sqrt{\gamma^2/4-\omega_q^2}} \sinh(t\sqrt{\gamma^2/4-\omega_q^2}\, )+ \theta(\omega_q^2-\frac{\gamma^2}{4} ) \,\frac{\frac{\gamma}{2}}{\sqrt{\omega_q^2-\gamma^2/4}} \sin(t\sqrt{\omega_q^2-\gamma^2/4} )
\end{eqnarray}
and $\theta(x)$ is the Heaviside function. The behavior of the time-dependent correlation strongly
depends on the value of the wavevector $q$: one can distinguish two types of regimes. For values of $q$
below the threshold $q_{thres}=2\arcsin(\frac{\gamma}{4\omega_E}) $ the correlations are overdamped, i.e. decay without oscillations (subcritical regime).
By contrast, for values of $q$ above the threshold, an underdamped oscillatory regime takes place (supercritical regime).
To observe the latter regime, it is necessary that the friction coefficient $\gamma$ is not too large compared with the
Einstein frequency, i.e. that the condition $(\frac{\gamma}{4\omega_E})\leq 1$ is fulfilled.
Underdamped systems respond with damped sinusoidal behavior, whereas overdamped responses have exponential responses and no oscillatory behavior.
The dynamics depend on different time scales: the persistence time $\tau$, the viscous time $\gamma^{-1}$ and the
wavevector dependent time $\omega_q^{-1}$. For small values of the wavevector $q$ in the subcritical regime, $\omega_q^{-1}$ represents the slowest time scale which determines the monotonic decay of $C_{uu}(q,t)$ as shown in Fig.~\ref{fig:Intermediate}~(a).
By contrast, above the threshold $q_{threshold}$ in the supercritical regime, the phonon correlation becomes underdamped, oscillates and the amplitude decreases as the inertial time $1/\gamma$, as represented in Fig.~\ref{fig:Intermediate}~(c).
In general (small $q$), the intermediate scattering function of entropons is qualitatively similar to that of phonons, i.e.\ it is characterized by a monotonic behavior which becomes slower only when $\tau$ is the slower time scale, as shown in Fig.~\ref{fig:Intermediate}~(a).
As reported in Fig.~\ref{fig:Intermediate}~(b), the larger $\tau$ the slower the time decay of $C_{vv}^{entropon}(q,t)$.
In the supercritical regime, entropons not only decay slower but are also characterized by the suppression of time oscillations typical of phonons in this case. Figure~\ref{fig:Intermediate}~(d) reveals that this effect is much stronger as $\tau$ increases.

Let us remark that the entropon correlation function~\eqref{eq:entropon} unlike the phonon correlation~\eqref{eq:phonon} contains a term which decays monotonically with the typical scale of the persistence time, $\tau$. As we have seen in section~\ref{omegacorrelation}, this term gives a non-dispersive central peak near $\omega=0$ in the frequency domain.
Instead,
the amplitude of the displacement thermal fluctuations varies in time as:
\begin{equation}
C_{uu}^{phonon}(q,t)=\frac{T}{m} \frac{ 1}{\omega_q^2} e^{-\gamma t/2 } \Bigl(
{\cal C}(q,t) + {\cal S}(q,t) \Bigr)
\end{equation}
whereas, after a simple calculation, one can show that the entropon modes vary as:
\begin{eqnarray}&&
C_{uu}^{entropon}(q,t)
=\frac{T_a}{m} \frac{ 1}{\omega_q^2} e^{-\gamma t/2 } \Bigl(
{\cal C}(q,t) + {\cal S}(q,t) \Bigr) - \tau^2 C_{vv}^{entropon}(q,t) \,.
\label{eq:entropon2}
\end{eqnarray}
Thus we may write the formula
\begin{equation}
C_{uu}^{total}(q,t)= \frac{T+T_a}{T}C_{uu}^{phonon}(q,t)- \tau^2 C_{vv}^{entropon}(q,t),
\label{totalvv}
\end{equation}
showing that the displacement fluctuations contain a first term
proportional to $(T+T_a)/\omega_q^2$, which diverges as $q\to 0$, i.e. in the infinite "volume" limit,
plus a finite negative correction proportional to the dynamic velocity correlations.
Roughly speaking, in the time domain the major difference between the phonon excitations and the entropon excitations stems
from the presence of a the term $e^{-t/\tau}$ in the latter (see Eq.~\eqref{eq:entropon}).
Moreover, we remark that in the limit of vanishing potential, the velocity correlation approaches one, corresponding to free inertial AOUP. In addition, in this limit, the system satisfies the following Green-Kubo relation~\cite{green1952markoff,kubo1966fluctuation}
\begin{equation}
\lim_{\omega_q\to 0}\int_0^\infty dt C^{total}_{vv}(q,t)=\frac{T}{m} \int_0^\infty dt e^{-\gamma t}+
\frac{T_a}{m} \frac{1}{1-\gamma^2 \tau^2 } \int_0^\infty dt
\Bigl(
e^{-\gamma t} -\gamma \tau e^{- t/\tau}
\Bigl)=\frac{T}{m\gamma} + \frac{T_a}{m\gamma} =D_t+D_a \,,
\end{equation}
which relates the spatial diffusion coefficient to the stationary velocity autocorrelation function.
The last relation provides the value of the thermal diffusion coefficient, $D_t=\frac{T}{m\gamma}$, and active diffusion coefficient, $D_a=v_0^2\tau$, of the model in the non-interacting limit.

 \subsection{Mean square displacement}
To further investigate the steady-state dynamical properties of the system, we study the mean-square displacement $\text{MSD}(t)$ of a single particle as a function of time. The $\text{MSD}(t)$ is defined as
\begin{equation}
\text{MSD}(t)=
\langle(u_n(t)-u_n(0))^2 \rangle=2 ( C^{total}_{uu}(0,0) -C^{total}_{uu}(0,t))\, .
\end{equation}
We study the two important limits, $t\to 0$ and $t\to \infty$.
To ascertain the short-time properties of the correlations we first consider the following
expansion:
\begin{equation}
\langle u_q(t) u_{-q}(0) \rangle =\sum_{n=0}^\infty (-1)^n\frac{t^{2n}}{(2n)!}
\langle u_{q}^{(n)}(0) u_{-q}^{(n)}(0) \rangle =
\langle u_q(t) u_{-q}(0) \rangle-\frac{t^2}{2} \langle v_q(0) v_{-q}(0) \rangle +\dots
\end{equation}
and find that the MSD,
for small values of the time argument, i.e. $t/\tau\ll1$ and $\gamma\tau\ll1$, evolves ballistically and not diffusively,
a characteristic of the inertial model:
\begin{equation}
\text{MSD}(t)
\approx t^2\int_{0}^\pi \frac{dq}{ \pi}\ \langle v_q(0) v_{-q}(0) \rangle =c_{vv}(0)\, t^2
=t^2\,\Bigl[
\frac{T}{m} +
\frac{\frac{T_a}{m}}{1+\tau\gamma} \frac{1}{\left(1+\lambda^2 \right)^{1/2}}\Bigr] \,,
\label{selfvelocity0}
\end{equation}
where we used formula~\eqref{cvvtempiuguali}.
Thus the short-time behavior of $\sigma^2_0(t)$ is determined by the equal-time velocity correlation function.
In the opposite limit, $t\to \infty$, we are not able to derive an exact formula for the MSD,
$\text{MSD}(t)$, but we provide analytically its qualitative behavior.
Since the velocity correlation function $C_{vv}(q,t)$ has its maximum at $t=0$ for any value of $q$ and
decays asymptotically to zero as $t\to \infty$, we focus attention on the
contribution stemming from the phonon-like term, $ \frac{T+T_a}{T}\,C_{uu}^{phonon}(q,t)$ in Eq.~\eqref{totalvv}:
\begin{equation}
\text{MSD}(t)\approx
2\frac{T+T_a}{m}\int_{0}^\pi \frac{dq}{\pi} \frac{ 1}{\omega_q^2}\Bigl[ 1-e^{-\gamma t/2 } \Bigl(
{\cal C}(q,t) + {\cal S}(q,t) \Bigr)\Bigr] \,.
\end{equation}
Since the small $q$ region, the one corresponding to overdamped behavior, gives the largest contribution to the integral we expand the integrand as follows:
\begin{equation}
e^{-\gamma t/2 } \Bigl(
{\cal C}(q,t) + {\cal S}(q,t) \Bigr) \approx e^{- \frac{\omega^2_q}{\gamma} t }
\end{equation}
and for large $t$ we obtain the following result (see appendix~\ref{besselI0}):
\begin{eqnarray}
\text{MSD}(t)\approx
2\frac{T+T_a}{m}
\int_{0}^\pi \frac{dq}{\pi}
\frac{1}{ \omega^2_q} \,\,\,
\Bigl[ 1-e^{- \frac{\omega^2_q}{\gamma} t } \Bigr] \approx \frac{2 a^2}{\sqrt { \pi}} \, \omega_E (\frac{t}{\gamma})^{1/2} \,.
\label{asymptopticMSD}
\end{eqnarray}
The fractional exponent $1/2$ is identical to that of single-file systems~\cite{yoshida1981dynamic, marchesoni2006subdiffusion, illien2013active, dolai2020universal, debnath2023structure, ikeda2023harmonic}, i.e. systems of particles in one dimension with excluded volume usually provided by interactions diverging at the origin.
This exponent implies a subdiffusive motion for long times which is not affected by the activity.


\section{Conclusions}
\label{SecConclusions}

In this paper, we have studied how active forces influence the fluctuations of a system of interacting particles by considering a one-dimensional crystal subject to inertial dynamics and thermal noise.
Compared with a similar study~\cite{caprini2020time}, here we consider an underdamped system which allows us to calculate the displacement-displacement correlations and to identify two contributions: an equilibrium-like term assimilable
to the phonons of an ordinary equilibrium solid embedded in a low-viscosity medium and a non-equilibrium term accounting for the entropy production of the system.
Whereas the first contribution is associated with the spatially uncorrelated velocity fluctuations of the particles, the second embodies the correlated behavior of their velocities.
We have established a relation between correlation and response functions, and entropy production and we have shown that the last quantity is entirely due to active fluctuations, i.e.\ the entropons.
At variance with previous work on chains consisting of active particles~\cite{gupta2021heat, singh2021crossover, santra2022activity}, here we have studied static and dynamical correlations which we have analytically predicted. Specifically, we have calculated the spatial dependence of the one-body density and the static structure factor, as well as the single-particle mean-square displacement.
In addition, we have found analytical expressions for the intermediate scattering functions and single-particle mean-square displacement.

The two-time correlations show the crossover from overdamped dynamics to a regime of underdamped propagating elastic waves
as a function of the wavelength of the fluctuation mode. While the former regime can be observed also in systems with large viscosity, the second regime can only be observed if the particles are massive and subject to a low viscosity. These modes correspond to propagating waves in the solid. We remark that since the form of the correlation functions in the frequency representation is independent of the lattice structure and dimensionality, which only appear through the equilibrium dispersion relation, our results also hold for two and three-dimensional solids. The same statement applies to the time-dependent correlations at fixed $q$. On the contrary, the real space correlations show a strong dependence on the dimensionality, and therefore the present findings cannot be extrapolated to dimensions higher than one.

The present work fills an existing gap by providing a study of the highly non-trivial microscopic correlations of a solid-like elastic active system.
Regarding the future applications of active solid-state physics, we believe that there will be important developments in the field
of bacterial biofilms which can be used to probe the mechanical behavior of elastic active matter.
In addition, while we have provided a suitable microscopic theory, we believe that our work paves the way towards the development of inertial active matter field theories~\cite{arold2020active, te2021jerky, te2023microscopic} capable of distinguishing between entropon and phonon contributions.

\section*{acknowlegments}
\noindent
This paper is dedicated to the memory of Luis Felipe Rull (1949-2022). 	\\\\
H.L. acknowledges support by the Deutsche Forschungsgemeinschaft (DFG) through the SPP 2265 under the grant number LO 418/25.
 
\section*{AUTHOR DECLARATIONS}


\noindent
The authors have no conflicts of interest to disclose.

\section{APPENDICES}
\appendix


\section{Definition of Fourier transform and spectral form of correlators}\label{app:spectral}

The time Fourier transform of the particle displacement, which is identified by the hat symbol, satisfies the following 
relations:
\begin{eqnarray}&&
\int_{-\infty}^\infty dt e^{i\omega t}
 u_n(t) =   \hat u_n(\omega)
\\&&
 \int_{-\infty}^\infty \frac{d\omega}{2\pi} e^{-i\omega t}
\hat  u_n(\omega)  =  u_n(t) \,.
\end{eqnarray}
Similar definitions hold for the other variables, whereas $\delta$-correlated white noises, such that
$
\langle \xi_n(t) \xi_m(t') \rangle=\delta(t-t')  \delta_{nm}
$,
satisfy the following relation in Fourier space
\begin{equation}
\langle \hat \xi_n(\omega) \hat \xi_m(\omega') \rangle=\int_{-\infty}^\infty dt e^{i\omega t} \int_{-\infty}^\infty dt' e^{i\omega t'}
\langle \xi_n(t) \xi_m(t') \rangle=\int_{-\infty}^\infty dt e^{i\omega t} \int_{-\infty}^\infty dt' e^{i\omega t'}\, \delta(t-t')  \delta_{nm}=2\pi\,\delta_{nm} \,\delta(\omega+\omega') \,,
\end{equation}
and
$
\int_{-\infty}^\infty dt e^{i\omega t}
\langle \xi_n(t) \xi_m(0) \rangle=   \delta_{nm}\,.
$
For a colored noise with exponential memory, the time correlator reads:
\begin{equation}
C_{ff}(t-t') \equiv \langle f_n^a(t) f_m^a(t') \rangle=\delta_{nm} F_0^2 e^{-|t-t'|/\tau} \,,
\end{equation}
and its Fourier transform is
\begin{equation}
\Phi_{ff}(\omega)=\int_{-\infty}^\infty dt e^{i\omega t}
\langle f_n^a(t) f_m^a(0) \rangle= 2 F_0^2 \frac{\tau}{1+\omega^2 \tau^2}\, \delta_{nm} \,.
\label{a40}
\end{equation}
Now, our convention is that $\Phi(\omega)$ is the  Fourier transform in the frequency domain $\omega$ of the time correlator $C(s)$.
By considering the inverse relation
$
C_{ff}(s)=
\int_{-\infty}^\infty \frac{d\omega}{2\pi} e^{-i\omega s}\, \Phi_{ff}(\omega)
$,
we obtain the following relation between  the average $\langle \hat f_n^a(\omega) \hat f_m^a(-\omega)\rangle$ 
and  $\Phi_{ff}(\omega)$:
\begin{equation}
 \langle \hat f_n^a(\omega) \hat f_m^a(\omega')\rangle=\Phi_{ff}(\omega) 2\pi \delta(\omega+\omega').
\end{equation}

\section{Trajectory calculation of the Entropy production rate of the medium}
\label{EPRapendix}
For completeness, we derive the formula for the entropy production rate used in the main text, i.e.\ Eq.~\eqref{eq:entropyprodrate}.
By using Eqs.~\eqref{dynamicequation0}-\eqref{dynamicequation2}, we can express the noise of the $n$-particles as a function of the system variables:
\begin{equation}
\xi_n= \frac{1}{ \sqrt{2T \gamma} }\Bigl(m\dot{v}_n +m\gamma v_n -F_n - f^a_n\Bigr) \,.
\label{xiev}
\end{equation}
The white noise $\xi_n(t)$ at time $t$ is governed by a Gaussian probability distribution with unit variance.
Since white noises are $\delta$-correlated, the probability distributions of $\xi_n$ at different times are independent.
As a consequence, the probability of noise path $\{\xi_n\}_{t_i}^{t_f}$ from time $t_i$ to the final time $t_f$ is given by
\begin{equation}
P[\{\xi_n\}_{t_i}^{t_f}]={\cal N} \exp\left(-\frac{1}{2} \int_{t_i}^{t_f} dt\, \xi_n^2(t)\right) \,,
\label{pxi}
\end{equation}
where ${\cal N} $ is a normalization constant.
Substituting Eq.~\eqref{xiev} into Eq.~\eqref{pxi} gives the path weight for a trajectory $v_n$ in the forward dynamics between an initial state at time $t_i$ and a final state at time $t_f$
\begin{equation}
\mathcal{P}_F[\{v_n\}_{t_i}^{t_f}]={\cal N}'\, \exp\Biggl(-\frac{1}{4T \gamma} \int_{t_i}^{t_f} dt \Bigl(m\dot{v}_n +m\gamma v_n-F_n - f^a_i\Bigr)^2 \Biggr) \,,
\end{equation}
where ${\cal N}'$ is another normalizer that also contains the Jacobian of the transformation from noise variables to dynamical variables (position, velocity, and activity).
From the path-probability probability, we can calculate the entropy production of the medium by applying its definition
\begin{equation}
S_m = \log\left( \frac{\mathcal{P}_F}{\mathcal{P}_R} \right)\,,
\end{equation}
where $\mathcal{P}_F$ is the probability of the forward trajectory of all the particles and $\mathcal{P}_R$ is the corresponding time-reversed probability.
Specifically, $\mathcal{P}_F$ reads:
\begin{equation}
\label{eq:app_forwardtraj}
\mathcal{P}_F \propto \prod_{n}^N\exp{\left(-\frac{1}{4T\gamma}\int_{t_i}^{t_f} dt\left( m \dot{v}_n +m\gamma v_n -F_n - f^a_n \right)^2\right)} \,,
\end{equation}
because the noise trajectories of different particles are independent.
On the other hand, the weight for the reverse path is found by applying the time-reversal transformation.
This means transforming time as $t\to-t$, the particle position as $x_n \to x_n$, and, consequently, the particle velocity as $v_n\to - v_n$.
Motivated by previous studies \cite{cagnetta2017large, shankar2018hidden, szamel2019stochastic, grandpre2021entropy, caprini2023entropons}, we assume that the activity $f^a_n$ is even under time-reversal transformation, such that $f^a_n \to f^a_n$. With this choice, even a potential-free AOUP particle in the presence of a thermal bath produces entropy, as expected in non-equilibrium systems.
Therefore, by setting $v_n\to -v_n$ in Eq.~\eqref{eq:app_forwardtraj}, we can easily calculate the probability of the reversed path $\mathcal{P}_R$ as
\begin{equation}
\mathcal{P}_R\propto \prod_{n}^N\exp{\left(-\frac{1}{4T\gamma}\int_{t_i}^{t_f} dt\left( m \dot{v}_n - m\gamma v_n-F_n - f^a_n \right)^2\right)} \,.
\end{equation}
We can therefore construct the ratio of the forward and backward path weights that carry us from the initial to the final state as:
\begin{equation}
\frac{\mathcal{P}_F}{\mathcal{P}_R}=\prod_{n}^N\exp{\left(-\frac{1}{4T\gamma}\int_{t_i}^{t_f}dt\left( m \dot{v}_n +m\gamma v_n
-F_n - f^a_n \right)^2\right)}\, \exp{\left(\frac{1}{4T\gamma}\int_{t_i}^{t_f}dt\left( m \dot{v}_n- m\gamma v_n-F_n - f^a_n \right)^2\right)}\,.
\end{equation}
With this definition, we have:
\begin{equation}
\begin{aligned}
S^m &=\sum_{n}^N \sigma_n^{m} \,,
\end{aligned}
\end{equation}
where $\sigma_n^m$ is the entropy production of the medium of the $n$-particle, that reads
\begin{equation}
\begin{aligned}
\sigma_n^m&=-\frac{1}{T\gamma} \int_{t_0}^{t_f} dt \left( m\dot{v}_n -F_n -f^a_n \right) \gamma v_n\\
&=-\frac{1}{T} \int_{t_0}^{t_f} dt \left( m\frac{d}{dt} \frac{v_n^2}{2} + v_n\nabla_{x_n} U - v_n f^a_n \right) \,.
\end{aligned}
\end{equation}
By summing over $n$, we have
\begin{equation}
S^m =\frac{1}{T}\left(\mathcal{K}_{t_0} - \mathcal{K}_{t_f}\right)
- \frac{1}{T} \sum_n \int_{t_0}^{t_f} dt v_n\left(\nabla_{x_n} U - f_n^a\right)\,,
\end{equation}
where $\mathcal{K}_{t_{\alpha}}=\sum_n mv_n^2/2$ is the total kinetic energy calculated at time $t_{\alpha}$.
If the force is an internal force we use
\begin{equation}
\begin{aligned}
\frac{1}{T} \sum_{nn'} \langle \,v_n \nabla_{x_n} U(x_n-x_{n'}) \rangle &=\frac{1}{2T} \sum_{nn'} \langle \,[v_n \nabla_{x_n} U(x_n-x_{n'}) +v_n \nabla_{x_{n'}} U(x_n-x_{n'})]\rangle \\
&= \frac{1}{2T} \frac{d}{dt} \sum_{nn'} U(x_n-x_{n'})=\frac{1}{T} U_{tot}\,.
\end{aligned}
\end{equation}
Thus, we may rewrite
\begin{equation}
S^m =\frac{1}{T}\left(\mathcal{K}_{t_0} - \mathcal{K}_{t_f}\right) +\frac{1}{T}\left(U_{t_0} - U_{t_f}\right)
+ \frac{1}{T}\sum_n \int_{t_0}^{t_f} dt \langle v_n f_n^a \rangle \,.
\end{equation}
If we consider the entropy production rate, we may divide the previous result by $(t_f - t_i)$ and observe that the first two terms
are boundary terms irrelevant for large times, $t_f - t_i$.
From here we obtain the expression for the entropy production rate of the medium
\begin{equation}
\dot S^m = \frac{1}{T} \sum_n \langle v_n f_n^a \rangle \,,
\end{equation}
which corresponds to Eq.~\eqref{eq:entropyprodrate}.

\section{Integrals over wavevectors}
\label{spatialintegrals}
The first integral in Eq.~\eqref{cuuintegral} can be computed as follows:
\begin{equation}
Q_0=\frac{(T+ T_a) }{m }\int_{2\pi/N}^{2\pi}\frac{ dq}{2\pi}\frac{1}{\omega_q^2} =
\frac{(T+ T_a) }{m \omega_E^2}\int_{2\pi/N}^{2\pi}\frac{ dq}{2\pi} \frac{1}{4\sin^2(q/2)} 
\approx \frac{(T+ T_a) }{m \omega_E^2}\frac{N}{2\pi^2}
\end{equation}
We consider now the first integral in Eq.~\eqref{twointegrals} 
\begin{equation}
Q_n= \frac{(T+ T_a) }{m \omega_E^2}
\int_{0}^{2\pi}  \frac{dq}{2\pi} \, \frac{ 1 - \cos(n q)    }{(1-\cos(q))}=4\frac{(T+ T_a) }{m \omega_E^2} \int_{0}^{\pi/2}  \frac{dx}{2\pi} \, \frac{\sin^2(n x)  }{\sin^2(x)}
\approx n\int_{0}^{n\pi/2}  dy \, \frac{\sin^2(y)  }{y^2} \,.
\end{equation}
Using the tabulated integral:
\begin{equation}
\int_{0}^{\infty}\frac{sin^2(\alpha x)}{x^2}=\alpha\frac{ \pi}{2}
\end{equation}
we find
\begin{equation}
Q_n\approx \frac{(T+ T_a) }{m \omega_E^2} n\,.
\end{equation}
The second  integral in the expression for the single-particle mean-square displacement, i.e.\ Eq.~\eqref{twointegrals} reads:
\begin{equation}
I_n=\frac{T_a}{m}\tau^2\int_{0}^\pi  \frac{dq}{\pi} 
\frac{ 1 - \cos(n q) }{1+\tau\gamma+\tau^2\omega_q^2}=
\frac{T_a}{m}\tau^2\frac{ 1 }{1+\tau\gamma}\int_{0}^\pi  \frac{dq}{\pi} \, \frac{ 1 - \cos(n q)  }{1+2\lambda^2(1-\cos(q))}
\end{equation}
and can be evaluated with the help of the tabulated result:
\begin{equation}
\int_0^\pi dx \frac{\cos(nx)}{1+\beta \cos(x)}=\frac{\pi}{\sqrt{1-\beta^2}}\Bigl( \frac{1-\sqrt{1-\beta^2}} {|\beta|}      \Bigr)^n
\end{equation}
with
\begin{equation}
\beta=-\frac{2\lambda^2}{1+2\lambda^2} 
\end{equation}
and $n\geq 0$. By using  the result of this integral, we obtain:
\begin{eqnarray}&&
I_n
=2\tau^2 \frac{T_a}{m}\, 
\frac{1}{1+\tau\gamma}  \frac{1}{\left(1+4\lambda^2 \right)^{1/2}}\, \Bigl[1-\Bigl( \frac{1+2\lambda^2-\sqrt{1+4\lambda^2}} {2\lambda^2}      \Bigr)^n  \Bigr] \,.
\end{eqnarray}
By considering the following approximation: 
\begin{equation}
\Bigl(    \frac{1+2\lambda^2-\sqrt{1+4\lambda^2}} {2\lambda^2}       \Bigr)^n    \approx \exp\left( n\ln(1-\frac{1}{\lambda})\right)\approx
\exp\left(-\frac{n}{\lambda} \right) \,,
\end{equation}
we finally obtain
\begin{eqnarray}&&
\langle (u_n-u_0)^2 \rangle
=na^2-2\tau^2 \frac{T_a}{m}\, 
\frac{1}{1+\tau\gamma}  \frac{1}{\left(1+4\lambda^2 \right)^{1/2}}\left(1-\exp{\left(-\frac{n}{\lambda}\right)}\right)\,,
\end{eqnarray}
which coincides with the result of Eq.~\eqref{resultintegrals}.

\section{Derivation of Eq.(\ref{asymptopticMSD})}
\label{besselI0}
Within the overdamped approximation valid for modes having $q<q_{thres}$ we write:
\begin{equation}
\text{MSD}(t)=\frac{T+T_a}{m \pi} \int_{-\pi}^\pi dq  \frac{1}{ 2\omega_E^2(1 -\cos(q))}
\left(1-e^{- 2 \omega_E^2(1-\cos(q) )t  /\gamma } \right) \,.
\end{equation}
To perform the integral, we consider its time derivative, which reads:
\begin{equation}
\frac{d}{dt}\text{MSD}(t)=
\frac{T+T_a}{m \gamma \pi}  e^{-(2\omega_E^2/\gamma) t} \int_{-\pi}^\pi dq \, e^{ 2 \frac{\omega_E^2}{\gamma} \cos(q) t   } = 2 \frac{T+T_a}{m \gamma }    e^{-(2\omega_E^2/\gamma) t}  I_0\left(\frac{2\omega_E^2}{\gamma}t\right)
\end{equation}
where $I_0(x)$ is the modified Bessel function of the first kind defined as
\begin{equation}
I_0(z)=\frac{1}{\pi}\int_0^\pi d\theta \, e^{z \cos\theta} \,. 
\end{equation}
By taking the asymptotic expansion for large $z$, we can write
\begin{equation}
I_0\left(\frac{2\omega_E^2}{\gamma}t\right)=\frac{e^{ (2\omega_E^2/\gamma) t }}{\sqrt{2\pi (2\omega_E^2/\gamma) t } } \left(1+\frac{1}{8}\frac{1}{(2\omega_E^2/\gamma) t }+\dots \right) \,.
\end{equation}
In the opposite limit of small $z$, the modified Bessel function can be approximated as $I_0(z)\approx (1+z/4+\dots)$, and we obtain
\begin{equation}
\frac{d}{dt} \text{MSD}(t)\approx
\frac{T+T_a}{m \omega_E\sqrt{\pi \gamma} } \, t^{-1/2} \,.
\end{equation}
Thus integrating with the initial condition $\sigma^2_0(0)=0$, we finally obtain
\begin{equation}
\text{MSD}(t)\approx 
\frac{2}{\sqrt \pi}\frac{T+T_a}{m \omega_E\sqrt{\gamma} } \, t^{1/2} \,,
\label{sigmazero}
\end{equation}
which corresponds to Eq.~\eqref{asymptopticMSD}.


\bibliographystyle{apsrev4-1}

\bibliography{JPA.bib}
\end{document}